  \documentclass[aps,prl,twocolumn,english,groupedaddress,showpacs,floatfix]{revtex4}

  \usepackage{amsmath,bbm}
  \usepackage{psfig,graphics,graphicx}
  \usepackage{babel}

\begin{document}

  \title{Scaling and universality in rock fracture}

  \author{J\"orn Davidsen}
  \email[]{davidsen@pks.mpg.de}
  \affiliation{British Antarctic Survey, High Cross, Madingley Road,
	Cambridge CB3 0ET, United Kingdom}
  \author{Sergei Stanchits}
  \affiliation{GeoForschungszentrumZentrum Potsdam, Department 3.2,
  Telegraphenberg D423, 14473 Potsdam, Germany}
  \author{Georg Dresen}
  \affiliation{GeoForschungszentrumZentrum Potsdam, Department 3.2,
  Telegraphenberg D423, 14473 Potsdam, Germany}

  \date{\today}

  \begin{abstract}

We present a detailed statistical analysis of acoustic emission time series
from laboratory rock fracture obtained from different experiments on different materials 
including acoustic emission controlled triaxial fracture and punch-through tests.
In all considered cases, the waiting time distribution can be described
by a unique scaling function indicating its universality. This scaling function 
is even indistinguishable from that for earthquakes suggesting its 
general validity for fracture processes independent of time, space and magnitude scales.

  \end{abstract}


  \maketitle


The fracture of materials is technologically of enormous interest due to its economic and human cost \cite{sornette}.
Despite the large amount of experimental data and the considerable efforts undertaken \cite{liebowitz}, many questions about fracture have not yet been answered. In particular, there is no comprehensive understanding of rupture phenomena but only a partial classification in restricted and relatively simple situations. 
For example, many material ruptures occur by a ``one crack'' mechanism and a lot of effort is being devoted to the understanding, detection and prevention of the nucleation of the crack \cite{maloy92,marder96,daguier96,fineberg99,buehler06,maloy06}. Exceptions to the ``one crack'' rupture mechanism are heterogeneous materials such as fiber composites, rocks, concrete under compression and materials with large distributed residual stresses. In these systems, failure may occur as the culmination of a progressive damage involving complex interactions between multiple defects and microcracks. 

In particular, acoustic emission (AE) due to microcrack growth precedes the macroscopic failure of rock samples under constant stress \cite{lockner77,yanagidani85} or constant strain rate loading \cite{mogi62,scholz68}. This is an example of the concept of ``multiple fracturing'' --- the coalescence of spontaneously occurring microcracks leading to a catastrophic failure --- which is thought be applicable to earthquakes as well \cite{turcotte,scholz,beeler04}. Due to this and the similarity in their statistical behavior, acoustic emissions can be considered analogous to earthquake sequences. The temporal \cite{hirata87,ojala04}, spatial \cite{hirata87a} and size distribution \cite{mogi62} of AE events follow a power law, just as it is commonly observed for earthquakes \cite{frohlich93,utsu95}. Such power-law scaling can be considered indicative of self-similarity in the AE and earthquake source process \cite{hirata87}.

The time evolution of AE and earthquake data also display considerable differences. Laboratory rock fracture is dominated by a large number of foreshocks while seismicity in the Earth's crust is characterized by an abundance of aftershocks \cite{main00a}. Here, we show that despite this difference the probability density function (PDF) for the time interval between successive events is the same in both cases if they are rescaled with the mean waiting time or equivalently with the mean rate of occurrence. In particular, the PDF for laboratory rock fracture neither depends on the specific experiment nor on the specific material. These observations strongly suggest a universal character of the waiting time distribution and self-similarity over a wide range of activity in rock fracture.


In Ref.~\cite{corral04} it was shown that the PDF of earthquake waiting times --- without distinguishing between foreshocks, main shocks or aftershocks --- for different spatial areas, time windows and magnitude ranges can be described by a unique distribution if time is rescaled with the mean rate of seismic occurrence. It was shown in particular that the distribution holds from worldwide to local scales, for quite different tectonic environments and for all magnitude ranges considered. This is even true if the seismic rate is not stationary as during aftershock sequences when the rate decays according to Omori's law \cite{omori}. In those cases, the waiting times have to be rescaled by the instantaneous rate instead \cite{corral04}.

Here, we analyze in the same way the waiting times between AE events during periods of \emph{stationary activity} in time series of laboratory rock fracture obtained from different experiments.
Laboratory experiments were performed on five different materials: Flechtingen sandstone (Fb, porosity about $7\%$), Bleuerswiller sandstone (Vo, porosity $24\%$), Aue granite (Ag, porosity $1.3\%$), Tono granite (To, porosity $1.7\%$) and Etna Basalt (Eb, porosity $2.1\%$). Sandstone samples were fractured at wet condition (pore pressure 10 MPa), granite and basalt samples at dry conditions. The formation of compaction bands was observed in the case of Vo sandstone and brittle fracture in all other cases. We investigated the fracture of rock samples at confining pressures in the range 5-100 MPa in different axial loading conditions: constant displacement rate of 20 $\mu$m/min (CDR), AE activity feedback control of loading (AFC), as described in Ref.~\cite{zang00}, and punch-through loading conditions (PT) \cite{backers02}. The threshold level of the rate control sensor allowed varying the speed of the fault propagation by three orders of magnitude, i.e. from mm/s in CDR tests to $\mu$m/s in AE AFC tests. The newly developed data acquisition system (DaxBox made by PR\"OKEL, Germany) records fully digitized waveforms (16 bit amplitude resolution, 10 MHz sampling rate) in 6~Gb memory buffer, providing zero dead time of registration (see \cite{stanchits06} for a detailed discussion). Most importantly, the data acquisition system allows to record AE events continuously even for high AE activity which is especially important for the analysis here. The only limitation on the shortest registered time intervals between subsequent AE events comes from the finite duration of the associated signals leading to the possibility of overlapping AE signals, for example, duplets or triplets. To simplify the fully automatic procedure of onset time picking and hypocenters location, only the first signal is located in a sliding window of 100 $\mu$s. Thus, the shortest time interval in the experiments considered here equals 100 $\mu$s, yet the onset times of AE arrivals at each particular channel were determined with accuracy of about 0.5 $\mu$s. 
For each experiment, we selected one or more periods of stationary activity for our analysis.

%
 \begin{table}
 \caption{\label{table}List of analyzed rock fracture experiments. Here, $P_c$ is the confining pressure, 
	$A_{th}$ is the selected AE amplitude threshold,
	$\langle T \rangle$ is the mean waiting time and $N$ the number of AE events.}
 \begin{ruledtabular}
 \begin{tabular}{llcccc}
	name & loading & $P_c$ (MPa) & $A_{th}$ (V) & $\langle T \rangle$ (sec) & $N$\\
	\hline
	FB38 & AFC & 50 & 0 & 0.0777 & 10339\\
	Vo1\_b & CDR & 100 & 1.0 & 0.0509 & 12817\\
	& & & 2.0 & 0.189 & 3447\\
	Vo2\_a & CDR & 60 & 0.0 & 0.110 & 12708\\
	Vo2\_b & CDR & 60 & 0.0 & 0.0687 & 26218\\
	Vo3\_c & CDR & 80 & 0.0 & 0.0453 & 30287\\
	& & & 0.6 & 0.0943 & 14564\\
	& & & 1.0 & 0.222 & 6188\\
	& & & 2.0 & 1.04 & 1318\\
	Ag72 & AFC & 20 & 0.0 & 0.154 & 4465\\
	Ag73 & AFC & 20 & 0.0 & 0.337 & 9860\\
	Ag74 & AFC & 10 & 0.0 & 0.120 & 10486\\
	& & & 1.0 & 0.453 & 2780\\
	Ag75\_a & AFC & 20 & 0.0 & 0.0479 & 1838\\
	To2\_2\_a & PT & 5 & 0.0 & 0.0371 & 1445\\
	To2\_3\_a & PT & 30 & 1.0 & 0.0364 & 1700\\
	To5\_2 & AFC & 20 & 0.0 & 0.230 & 21135\\
	& & & 0.6 & 1.07 & 4538\\
	& & & 1.0 & 2.58 & 1889\\
	To5\_3 & AFC & 30 & 0.0 & 0.369 & 2263\\
	Eb12 & AFC & 20 & 0.0 & 0.776 & 1632\\
 \end{tabular}
 \end{ruledtabular}
 \end{table}

In the following, we focus on two quantities to characterize each AE event: time of occurrence and AE adjusted amplitude calculated according to the procedure described in Ref.~\cite{zang98}. While in most cases we consider all recorded AE events, we also study the effect of detection thresholds by only considering events with AE adjusted amplitude $A$ above a certain threshold $A_{th}$. In both cases, the AE series is transformed into a point process where events occur at times $t_i$ with $1 \le i \le N$, and therefore, the time between successive events can be obtained as $T_i = t_{i+1} - t_i$. These are the waiting times which are also referred to as recurrence times or interoccurrence times. For a given fracture experiment $E$, their PDF is denoted by $P_E(T)$.

\begin{figure}
   \includegraphics*[width=\columnwidth]{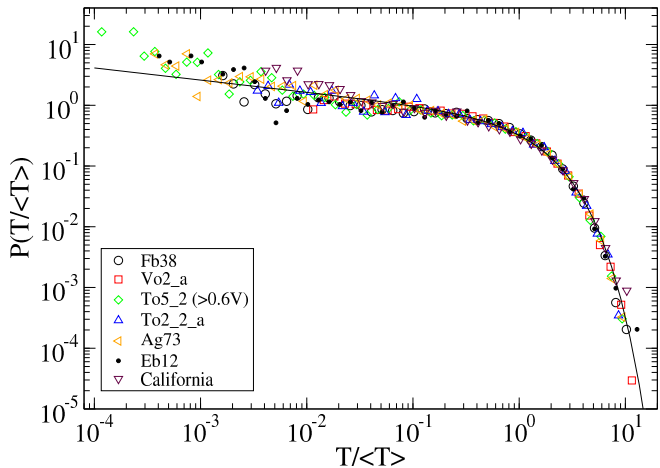}
   \caption{\label{all}
   (Color online) Probability density function of normalized waiting times $T/\langle T \rangle$ 
	for five different rock fracture
	experiments (see Table~\ref{table} for details) and an earthquake catalog from Southern 
	California for comparison (see text for details).
	The solid line corresponds to a fit based on Eq.~(\ref{eq_wait}) with $\theta = T/\langle T \rangle$ giving $\gamma \approx 0.8$ and $B \approx 1.4$.}
   \end{figure}

Fig.~\ref{all} shows the PDF of the normalized waiting times $\theta = T/\langle T \rangle$ for different rock fracture experiments where $\langle T \rangle$ is the respective mean waiting time. The excellent data collapse implies that $P(T/\langle T \rangle)$ does not depend on the particular rock fracture experiment and that we can write
\begin{equation}
	P_E(T) = P(T/\langle T \rangle_E) / \langle T \rangle_E.
\end{equation}
Thus for a given fracture experiment $E$, the PDF $P_E(T)$ of its AE series is determined by its mean waiting time $\langle T \rangle_E = (t_N - t_1)/(N - 1)$ --- or equivalently the mean rate $R_E = \langle T \rangle_E^{-1}$ --- and the universal scaling function $P(\theta)$ which can be well approximated by a Gamma distribution
\begin{equation} \label{eq_wait}
	P(\theta) \propto \theta^{-(1-\gamma)} \; \exp{(-\theta/B)},
\end{equation}
with $\gamma \approx 0.8$, $B \approx 1.4$ and the prefactor fixed by normalization \cite{note1}.
Therefore, we have essentially a decreasing power-law with exponent about 0.2, up to the largest values of the argument, $\theta = T/\langle T \rangle$ about 1, where the exponential factor comes into play. This is statistically indistinguishable (at the 2$\sigma$ level) from the results for earthquake data given in Ref.~\cite{corral04}, namely $\gamma = 0.67 \pm 0.05$ and $B = 1.58 \pm 0.15$. As shown in Fig.~\ref{all}, this is also confirmed by the PDF of an earthquake series from Southern California which we have included for comparison \cite{note2}.

\begin{figure}
   \includegraphics*[width=\columnwidth]{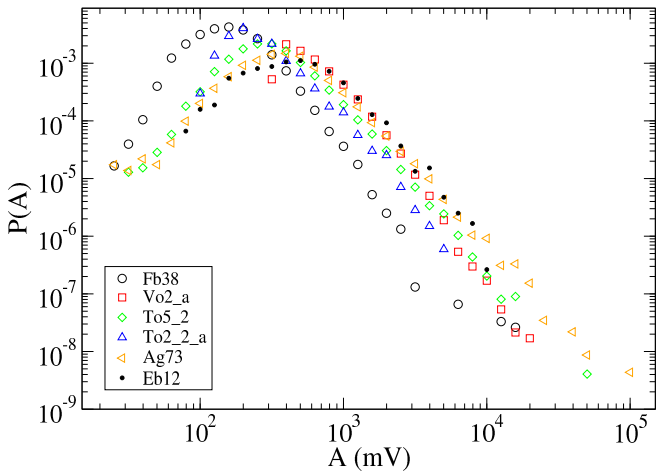}
   \caption{\label{GR}
   (Color online) Probability density function of AE adjusted amplitude $A$ for the same rock 
	fracture experiments as in Fig.~\ref{all}.}
   \end{figure}

In particular, $P(\theta)$ is independent of the frequency-amplitude distribution of the AE signal. As Fig.~\ref{GR} shows, the PDF $P(A)$ for the AE adjusted amplitude $A$ depends crucially on the particular experiment. While the overall structure of $P(A)$ is rather similar for the different experiments --- a sharp increase up to a maximum value followed by a power-law like decrease --- details such as the location of the maximum and the slope $\beta$ of the tail can be very different. For instance, the latter varies between 2.5 and 3.7 for the considered experiments. This corresponds to a variation in the Gutenberg-Richter exponent $b$ between 1.1 and 1.9 which characterizes the frequency-magnitude relation of earthquakes \cite{gutenberg,note3}. 
Thus for our rock fracture experiments, the variation in $b$ is larger than the regional variability in the earthquake data studied in \cite{corral04}. Yet, $P(\theta)$ remains basically unchanged. More importantly, even considering only events above a lower threshold $A_{th}$, as for the data set To5\_2 in Fig.~\ref{all}, does not affect $P(\theta)$. This indicates the robustness of our results.

\begin{figure*}[htbp]
   \includegraphics*[width=\textwidth]{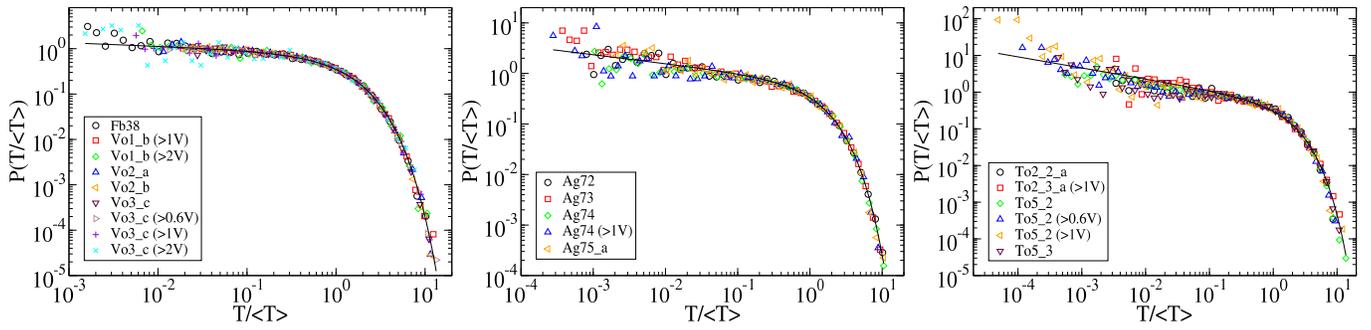}
   \caption{\label{comb}
   (Color online) Probability density functions of normalized waiting times $T/\langle T \rangle$ 
	for fracture experiments on sandstone, Aue granite and Tono granite (from left to right;
	see Table~\ref{table} for details).
	The solid lines correspond to best fits according to Eq.~(\ref{eq_wait}) giving 
	$\gamma = 0.92$ and $B = 1.2$, $\gamma = 0.82$ and $B = 1.3$, $\gamma = 0.70$ and $B = 1.5$, respectively.}
   \end{figure*}

This robustness is further confirmed by Fig.~\ref{comb} which shows $P(\theta)$ for a large selection of fracture experiments (see Table~\ref{table} for details). Again, $P(\theta)$ can be well described by Eq.~(\ref{eq_wait}). The slight variation in the fitted values of $\gamma$ and $B$ can be attributed mainly to statistical fluctuations and partially to measurement induced biases: The relatively high value of $\gamma$ for sandstone is a consequence of the inability to detect the shortest waiting times due to measurement restrictions absent in the other experiments. This absence of short waiting times (an order of magnitude compared with granite) significantly biases the estimate of $\gamma$ towards higher values.

Fig.~\ref{comb} shows not only that for sandstone and different types of granite the influence of the specific material on $P(\theta)$ is neglectable but also that the type of experiment (punch-through vs. constant displacement rate vs. activity feedback control) has no significant influence on $P(\theta)$. Moreover, Fig.~\ref{comb} indicates that variations with $A_{th}$ are neglectable as well. Even restricting the included AE events to arbitrarily selected areas within the rock sample did not alter $P(\theta)$ (not shown). 
All these observations strongly suggest that $P(\theta)$ given in Eq.~(\ref{eq_wait}) is a universal result for rock fracture. It further implies that $P(T)$ is self-similar over a wide range of activity rates spanning two orders of magnitude for the experiments considered here alone (see Table~\ref{table}).

Our results also indicate that the universal form of $P(\theta)$ can be recovered for AE signals with largely varying AE rates, as for example during foreshock sequences, if instantaneous rates are used. As Table~\ref{table} shows, the AE signal of experiment Vo2 consists of at least two long stationary regimes, Vo2\_a and Vo2\_b, with different $\langle T \rangle$'s. Yet, the respective PDFs $P(\theta)$ are indistinguishable as follows from Fig.~\ref{comb}. This implies that $P(\theta)$ for the combined signal is the same as well \cite{note4}.

While we have presented strong evidence that $P(\theta)$ is universal for AE signals in rock fracture and earthquake sequences, the correlations between subsequent waiting times are very different. In Ref.~\cite{livina05}, it was shown that the distribution of waiting times between earthquakes strongly depends on the previous waiting time, such that small and large waiting times tend to cluster in time. We find that this is not the case for the AE signals studied here. In contrast, the conditional PDF $P(\theta|\theta_0)$ is independent of the previous waiting time $T_0$ with $\theta_0 = T_0 / \langle T \rangle$ and, thus, $P(\theta|\theta_0) = P(\theta)$. This might be due to the small number of pronounced foreshock and aftershock clusters of which the latter are particularly dominant in seismicity. 


To summarize, we have shown that the probability density function for waiting times in laboratory rock fracture is self-similar with respect to the AE rate and can be described by a unique and universal scaling function $P(\theta)$. Its particular form can be well approximated by a Gamma function implying a broad distribution of waiting times. This is very different from a narrow Poisson distribution expected for simple random processes and indicates the existence of a non-trivial universal mechanism in the AE generation process. The similarity with seismicity even suggests a connection with fracture phenomena at much larger scales and might help to understand this mechanism. 

JD would like to thank C. Goltz for stimulating discussions.


\begin{thebibliography}{30}
\bibitem{note1} The deviations between the data and Eq.~(\ref{eq_wait}) for small arguments (see Fig.~\ref{all}) are partially due to weak non-stationarities, namely foreshock activity, and partially due to statistical uncertainties. The statistical uncertainties are especially pronounced for small arguments since the number of events in the corresponding bins is low.
\bibitem{note3} We used $b = 1.44 \frac{\beta - 1}{2}$ which assumes $A^2 \propto E$ where $E$ is the total energy of an AE event. In the case of earthquakes, $E$ corresponds to the total energy in the seismic waves which can be related to the event's magnitude $m$ by $\log_{10}{E} = 1.44 \; m + 5.24$ \cite{turcotte}.
\bibitem{note4} If instantaneous rates are not used here, the non-stationarities lead to pronounced deviations at small and large arguments of $P(\theta)$. In particular, foreshock sequences with rate exponent $p'$ will lead to a faster decay with an expected exponent $2-1/p'$ at short time scales similar to what has been observed for aftershock sequences in seismicity \cite{corral04,shcherbakov05}.
\end{thebibliography}

  \end{document}